\documentclass[%
aps, %
prl, floatfix, 
twocolumn, 10pt, 
superscriptaddress, %
a4paper, %
amsfonts, amsmath, amssymb,longbibliography]{revtex4-2}
\usepackage[sort&compress]{natbib}
\usepackage{newtxtext}
\usepackage[varg]{newtxmath}
\usepackage{ascmac}
\usepackage{mathrsfs}
\usepackage{bm}
\usepackage{tabularx}
\newcolumntype{C}{>{\centering\arraybackslash}X}
\usepackage{color}
\usepackage{graphicx}
\usepackage{bm}
\usepackage{braket}
\usepackage{mathrsfs}
\usepackage[normalem]{ulem}
\usepackage{hhline}
\usepackage{hyperref}
\hypersetup{%
    unicode=false, %
    bookmarksnumbered=true, %
    bookmarkstype=toc, %
    breaklinks=true, %
    colorlinks=true, %
    pdfborder={0 0 1}, %
    linkcolor=blue, %
    citecolor=blue, %
    urlcolor=blue, %
    pdftoolbar=true, %
    pdfmenubar=true, %
    pdffitwindow=false, %
    pdfstartview={FitH} %
}
\newcommand{\bk}{\bm{k}}
\newcommand{\bp}{\bm{p}}

\newcommand{\zi}{i}
\newcommand{\tr}{\mathrm{tr}\,}

\newcommand{\hole}{h}
\newcommand{\eI}{e_1}
\newcommand{\eII}{e_2}
\newcommand{\eIII}{e_3}
\begin{document}
%
%
%
%
%
\title{Magnetic Moment vs Angular Momentum: Spin Hall Response in Bismuth}
\date{\today}
\author{Junji Fujimoto}
\affiliation{Department of Electrical Engineering, Electronics, and Applied Physics, Saitama University, Saitama, 338-8570, Japan}
\email[E-mail address: ]{jfujimoto@mail.saitama-u.ac.jp}

\author{Yuki Izaki}
\affiliation{Department of Physics, Tokyo Institute of Technology, 2-12-1 Ookayama, Meguro-ku, Tokyo 152-8551, Japan}

\author{Yuki Fuseya}
\affiliation{Department of Physics, Kobe University, Kobe 657-8501, Japan}

\begin{abstract}
Spin currents can carry either spin angular momentum or its associated magnetic moment, which are no longer strictly proportional in multiband systems.
Using a multiband $\bk\!\cdot\!\bp$ model, we compute the intrinsic spin Hall conductivity tensors of elemental Bi.
The magnetic-moment tensor emerges about two orders of magnitude larger and far less anisotropic than the angular-momentum tensor, while quasiparticle damping activates otherwise longitudinal components.
The magnetic-moment spin Hall angle exceeds unity, demonstrating that a clear distinction between the two currents is indispensable for multiband systems.
\end{abstract}
\maketitle
Spin currents can carry either the spin angular momentum~(SAM) or the spin magnetic moment~(SMM).
In a one-band electron system, these two operators are proportional, so they need not be distinguished.
In multiband systems, however, this proportionality is lost and the two operators must be treated separately.

This fact is already evident in the minimal example of the vacuum three-dimensional (3D) Dirac electron system.
The 3D Dirac system comprises two bands, each doubly degenerate in spin, and is governed by a $4 \times 4$ Dirac Hamiltonian.
Introducing Pauli matrices $\sigma^{\alpha}$ ($\alpha = x,y,z$) for the spin degeneracy, we define SAM, $\bm{s}=(s^{x}, s^{y}, s^{z})$, and SMM, $\bm{m}=(m^{x}, m^{y}, m^{z})$, as
\begin{align}
s^{\alpha}
    & = \frac{\hbar}{2}
        \begin{pmatrix}
            \sigma^{\alpha}
        &   0
        \\  0
        &   \sigma^{\alpha}
      \end{pmatrix}
, &
m^{\alpha}
    & = - \frac{g \mu_{\mathrm{B}} }{2}
        \begin{pmatrix}
            \sigma^{\alpha}
        &   0
        \\  0
        &   - \sigma^{\alpha}
      \end{pmatrix}
,\end{align}
where $\mu_{\mathrm{B}}$ is the Bohr magneton and $g$ is the $g$-factor~\cite{foldy1950}.
Because the two operators differ in sign structure, their responses to external perturbations can be markedly different.

This distinction becomes crucial in the context of the spin Hall effect.
In multiband materials, the choice of spin operator, SAM or SMM, can alter the predicted spin Hall conductivity (SHC) and, consequently, the interpretation of experiments.
SAM would couple to mechanical rotation~\cite{hehl1990,matsuo2011}, whereas SMM couples to magnetic fields~\cite{rajagopal1973,macdonald1979}.  
As a result, the magnitude of spin-transfer torque~\cite{slonczewski1996,berger1996} and the magneto-optical response~\cite{argyres1955} depend on which type of spin current flows and accumulates at a boundary.  
Resolving this distinction is therefore far from a mere technicality; it is essential for a correct description of spin-related physics and for an accurate understanding of phenomena such as the spin Hall effect, spin torques, and related topological responses.

In this Letter, we demonstrate that the choice of spin operator qualitatively alters the spin Hall response: within the minimal 3D Dirac model, the SHC associated with SAM vanishes identically, whereas that associated with SMM remains finite~\cite{fuseya2012,fuseya2014,fukazawa2017}.
We next consider elemental bismuth, a prototypical multiband semimetal whose electron and hole pockets reside at the $L$ and $T$ points; carriers near $L$ are described by an anisotropic 3D Dirac Hamiltonian~\cite{cohen1960,wolff1964,fuseya2014}.
The calculation of the SMM in multiband systems has long remained a formidable challenge, with analytical progress essentially limited to the two-band (Dirac) model. This fundamental difficulty has been a major impediment to exploring spin Hall effects driven by SMM. Recent theoretical advances based on $\bk\!\cdot\!\bp$ theory, however, have succeeded in overcoming this limitation~\cite{fuseya2015}, thereby opening the door to a systematic treatment of SMM in generic multiband systems.
Starting from the tight-binding model of Liu and Allen~\cite{liu1995} and employing a $\bk\!\cdot\!\bp$ theory, we construct both spin operators, the SMM via the general formula of the $g$-factor~\cite{fuseya2015}, and the SAM by projecting Pauli matrices onto the orbital basis at the $L$ and $T$ points, and compute the complete intrinsic SHC tensor.
Earlier theoretical studies of elemental bismuth addressed only SAM currents~\cite{sahin2015,guo2022,qu2023}.
While Ref.~\cite{chi2022} experimentally reported the magnitude of the SMM-SHC in polycrystalline bismuth, our work resolves its full crystal tensor and reveals an unexpected anisotropy reduction.

We present the first full-tensor evaluation of the SMM-based SHC in elemental bismuth, obtained on the same footing as its SAM counterpart.
Our complete SMM- and SAM-based SHC tensors for bismuth satisfy all crystalline-symmetry constraints~\cite{guo2022}.
The SMM-based SHC is substantially less anisotropic at the Fermi level than the SAM-based one.
By evaluating the electrical conductivity tensor and fitting two damping parameters, we reproduce the measured resistivities of bismuth~\cite{michenaud1972}, indicating the accuracy of our calculation.
The spin Hall angle derived from the SMM-based SHC exceeds unity, whereas that derived from the SAM-based SHC is two orders of magnitude smaller.
These findings demonstrate that a clear distinction between SAM and SMM is essential for an accurate description of the spin Hall effect.  
Because the underlying $\bk\!\cdot\!\bp$ theory requires only the low-energy band structure; the approach is readily extensible to narrow-gap semiconductors, Dirac electrons, and other topological materials in which multiband effects are crucial.

In the 3D Dirac model, the spin current operator that describes the flow of SAM is $j_{\mathrm{a}, i}^{\alpha} = (\hbar/2) \{ v_i, s^{\alpha} \} / 2$, where $v_i
    = c \begin{pmatrix}
        0
    &   \sigma^i
    \\  \sigma^i
    &   0
\end{pmatrix}$ is the velocity operator with the speed of light $c$, and $\{A, B\} = AB + BA$ is the anticommutator.
The corresponding operator for the flow of SMM is $j_{\mathrm{m}, i}^{\alpha} = \{ v_i, m^{\alpha} \} / 2$.
Evaluating the two operators yields
\begin{align}
j_{\mathrm{a}, i}^{\alpha}
    \propto
    \begin{pmatrix}
        0
    &   \sigma^0
    \\  \sigma^0
    &   0
    \end{pmatrix}
    \delta^{i \alpha}
, \quad
j_{\mathrm{m}, i}^{\alpha}
    \propto
    \begin{pmatrix}
        0
    &   \sigma^{j}
    \\  - \sigma^{j}
    &   0
    \end{pmatrix}
    \epsilon^{i \alpha j}
,\end{align}
where $\sigma^0$ is the $2 \times 2$ unit matrix, $\delta^{i \alpha}$ is the Kronecker delta, and $\epsilon^{i \alpha j}$ is the Levi-Civita symbol.
Thus, only the diagonal component ($i = \alpha$) of the SAM current survives at the operator level, whereas for the SMM current only off-diagonal components ($i \neq \alpha$) remain finite.
Within the linear response theory, $j_{\mathrm{a}, i}^{\alpha} = \sigma_{\mathrm{a}, i j}^{\alpha} E_j$, where $E_j$ is an applied electric field, it immediately follows that $\sigma_{\mathrm{a}, ij}^{\alpha} = 0$ for $i \neq \alpha$.  
Consequently, the 3D Dirac model provides a minimal setting in which the spin Hall response carried by SAM vanishes identically, whereas that carried by SMM is allowed, highlighting the essential difference between the two operators of spin in multiband systems.
This point has been pointed out by one of the present authors~\cite{chi2022}, and SMM-based spin Hall conductivity is evaluated in Refs.~\cite{fuseya2012,fuseya2014,fukazawa2017}.

\begin{figure}
\centering
\includegraphics[width=\linewidth,clip]{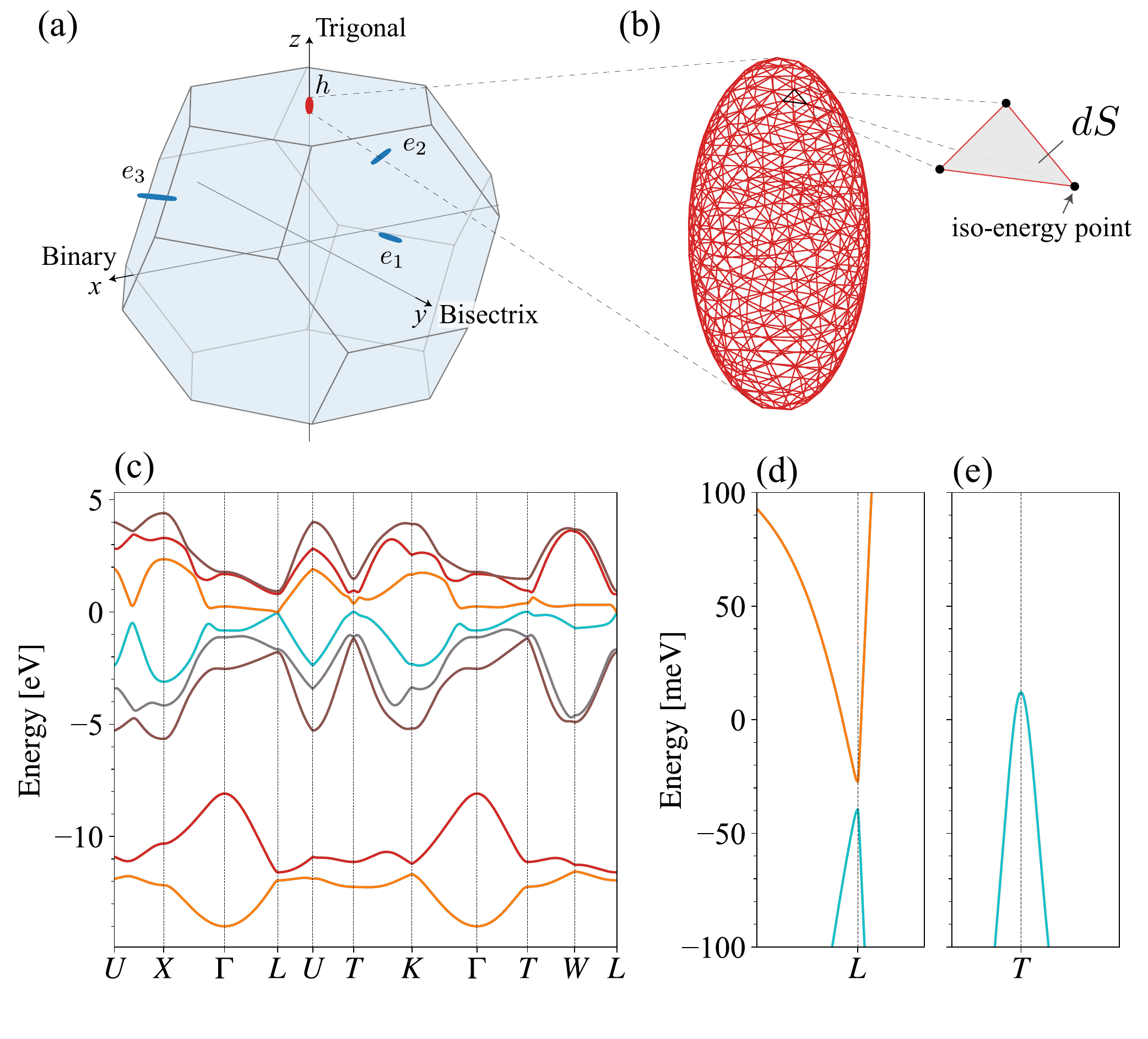}
\caption{\label{fig:bz-fs}
(a)~Brillouin zone of bismuth showing the hole pocket at the $T$ point (labelled $\hole$) and the three electron pockets at the symmetry-equivalent $L$ points (labelled $\eI$, $\eII$, and $\eIII$).
(b)~Triangulated representation of the hole-pocket Fermi surface; a representative triangular patch with area $dS$ and its three isoenergetic vertices are indicated.
(c)~Electronic band structure of bismuth, and energy dispersions near (d) the $L$ point and (e) the $T$ point, calculated with the Liu-Allen tight-binding model~\cite{liu1995}.
}
\end{figure}
We now evaluate the intrinsic spin Hall conductivities (SHCs) of bismuth.
Because the Fermi surfaces of bismuth are very small, see Fig.~\ref{fig:bz-fs}, the intrinsic contribution, governed by the spin Berry curvature, is expected to dominate, whereas the extrinsic contribution is expected to be negligible.
The small Fermi pockets also complicate numerical evaluations of transport coefficients; the $\bk\!\cdot\!\bp$ theory is well-suited to low-energy excitations and therefore offers an efficient and accurate approach.

The calculation procedures have been done as follows:
the effective Hamiltonians for the hole pocket ($h$) at the $T$ point and the three electron pockets ($e_1, e_2, e_3$) at the $L$ points, as shown in Fig.~\ref{fig:bz-fs}~(a), based on the $\bk\!\cdot\!\bp$ method for the tight-binding model by Liu-Allen~\cite{liu1995} are given as
\begin{align}
H_{\bk \cdot \bp}^{(\tau)}
    & = H_0^{(\tau)}
        + \sum_{\substack{i = x, y, z \\ n, m}} \hbar \left( v^{(\tau)}_i \right)_{n m} k_i \ket{\tau, n} \bra{\tau, m}
\end{align}
with the valley index $\tau = \hole, \eI, \eII, \eIII$, where $H_0^{(\tau)}$ is the diagonal matrix providing the eigenenergy for each band in the $\tau$ valley, $v^{(\tau)}_i$ ($i = x, y, z$) is the velocity operator, $\bk = (k_x, k_y, k_z)$ is the wavevector measured by the corresponding point, and $\ket{\tau, n}$ is the $n$-th eigenstate in the $\tau$ valley.
We retain two bands near the Fermi level for the $\eI$, $\eII$, and $\eIII$ valleys, and six bands for the $\hole$ valley (see Supplementary Material~(SM) for details).
The operator of the SMM, $m^{(\tau), \alpha}$, is evaluated through the general $g$-factor formula derived by $\bk\!\cdot\!\bp$ theory with the L\"owdin partitioning~\cite{fuseya2015}. 
The SAM operator $s^{\alpha}$ is defined as the spins associated with the atomic orbitals and expanded by the $\bk\!\cdot\!\bp$ method by using the same eigenstates for the SMM, which leads to $s^{(\tau), \alpha}$.
From here, we do not denote the valley index $\tau$ for readability unless it needs to be emphasized.
The corresponding spin current operators are given by $j_{\mathrm{m}, i}^{\alpha} = \{ v_i, m^{\alpha} \} / 2$ and $j_{\mathrm{a}, i}^{\alpha} = (\hbar/2) \{ v_i, s^{\alpha} \} / 2$, respectively, with $i, j, \alpha \in \{ x, y, z \}$.
See SM for more details of the definitions and derivations of SMM and SAM.
We numerically evaluate the electrical conductivity, $\sigma_{i j}$, and spin conductivities, $\sigma_{\mathrm{m}, i j}^{\alpha}$ and $\sigma_{\mathrm{a}, i j}^{\alpha}$, based on the Kubo formula;
\begin{align}
\sigma_{q, i j}^{\alpha}
	& = \lim_{\omega \to 0} \frac{K_{q, i j}^{\alpha} (\omega) - K_{q, i j}^{\alpha} (0)}{\zi \omega}
\end{align}
with $q = \mathrm{m}, \mathrm{a}$ and $K_{i j}^{\alpha} (\omega) = \mathcal{K}_{i j}^{\alpha} (\hbar \omega + \zi 0)$,
\begin{align}
\mathcal{K}_{i j}^{\alpha} (\zi \omega_{\lambda})
	& = \sum_{\tau} \frac{e}{\beta V} \sum_{n, \bk} \tr \left[
		j_{q, i}^{(\tau), \alpha}
		\hat{G}_{\bk +}^{(\tau)}
		\hat{v}_j^{(\tau)}
		\hat{G}_{\bk}^{(\tau)}
	\right]
\label{eq:sigma_q}
,\end{align}
where $G_{\bk}^{(\tau)} = G_{\bk}^{(\tau)} (\zi \epsilon_n) = (\zi \epsilon_n - H_{\bk\cdot\bp}^{(\tau)})^{-1}$ is the Green function, and $G_{\bk +}^{(\tau)} = G_{\bk}^{(\tau)} (\zi \epsilon_n + \zi \omega_{\lambda})$.
The electrical conductivity $\sigma_{ij}$ is obtained by replacing $j_{q, i}^{(\tau), \alpha}$ with $- e v_i^{(\tau)}$ in Eq.~(\ref{eq:sigma_q}).
We employed the relaxation time approximation (i.e., the constant damping rate approximation) to calculate the electrical conductivity and SHCs.
To fit the magnitude and anisotropy of the electrical conductivity at $T = 300~\mathrm{K}$~\cite{michenaud1972}, we set the damping rates as $\gamma_{\hole} = 0.07~\mathrm{meV}$ for $\hole$ and $\gamma_e = 0.16~\mathrm{meV}$ for $\eI, \eII, \eIII$.
In the $\bk$-integrals of the conductivities, we first integrate over the iso-energy two-dimensional surfaces~(see Fig.~\ref{fig:bz-fs}~(b)), and then integrate along the energy direction.
The Computational Geometry Algorithms Library~(CGAL)~\cite{cgal:ry-smg-24a} triangulates the iso-energy surfaces.
To compare the spin conductivities $\sigma_{\mathrm{m}, i j}^{\alpha}$ and $\sigma_{\mathrm{a}, i j}^{\alpha}$, we normalize them to be in the conductivity units~$(\mathrm{/\Omega \, m})$;
\begin{align}
\tilde{\sigma}_{\mathrm{m}, i j}^{\alpha} = \frac{e}{\mu_{\mathrm{B}}} \sigma_{\mathrm{m}, i j}^{\alpha}
, \qquad
\tilde{\sigma}_{\mathrm{a}, i j}^{\alpha} = \frac{2 e}{\hbar} \sigma_{\mathrm{a}, i j}^{\alpha}
.\end{align}
We introduce the energy cutoff for the validity of the $\bk\!\cdot\!\bp$ theory as $|E_{\mathrm{cutoff}} - E_{\mathrm{F}}| = 80~\mathrm{meV}$, where $E_{\mathrm{F}}$ is the Fermi energy, and set it to zero.

\begin{figure}
\centering
\includegraphics[width=\linewidth,clip]{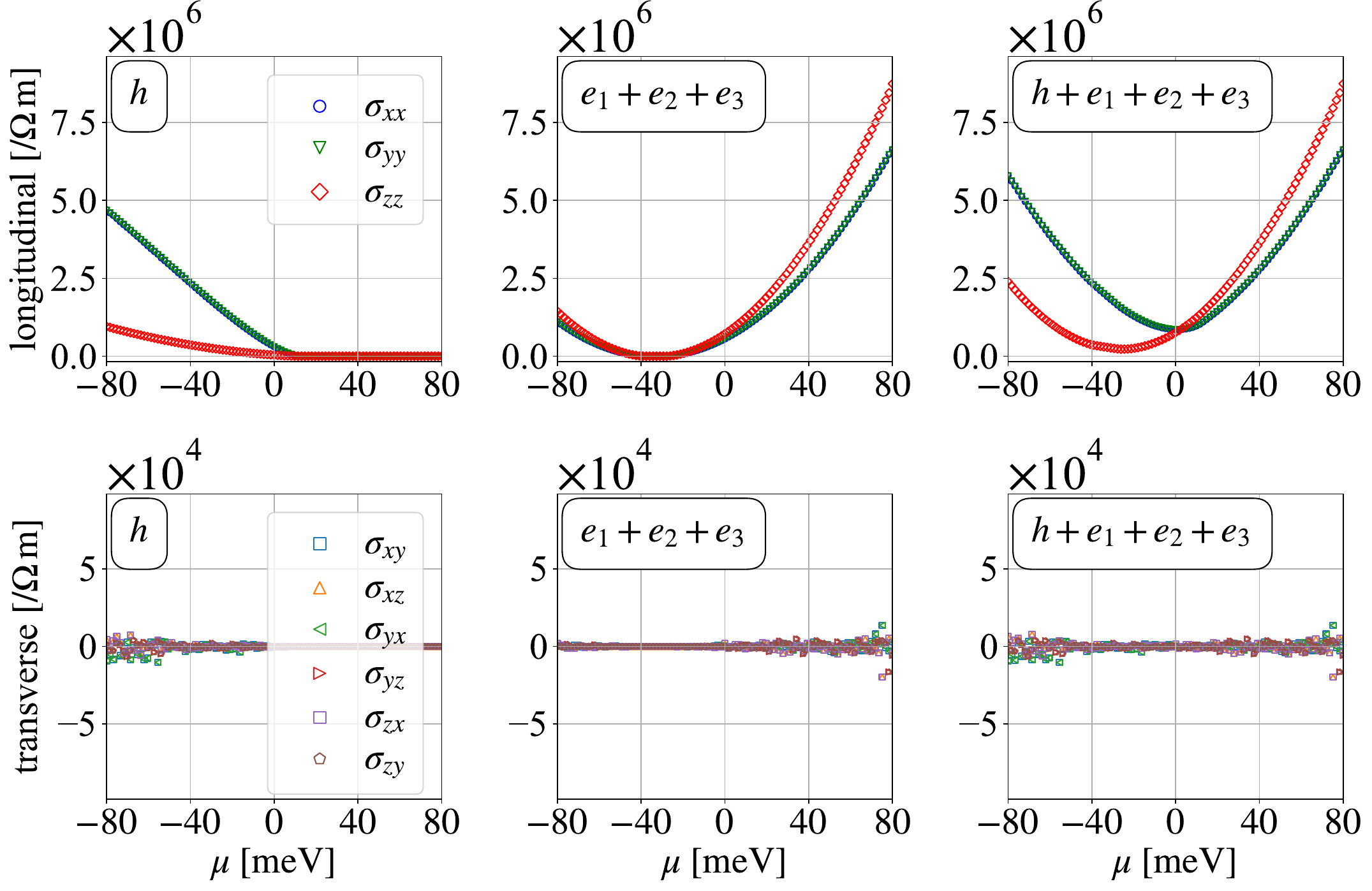}
\caption{\label{fig:electrical_conductivity_tensor}
Chemical-potential dependence of the electrical conductivity tensor for (left panels) the hole pocket at the $T$ point ($\hole$), (center panels) the combined electron pockets at the three symmetry‐equivalent $L$ points ($\eI$, $\eII$, and $\eIII$), and (right panels) their total.
The anisotropy of the total tensor respects the crystalline symmetry, whereas each individual electron pocket exhibits non-zero off-diagonal components (see the SM for details).
To reproduce the measured magnitude and anisotropy at $T = 300~\mathrm{K}$~\cite{michenaud1972}, we set the damping rates to $\gamma_{\hole} = 0.07~\mathrm{meV}$ for $\hole$ and $\gamma_{e} = 0.16~\mathrm{meV}$ for $\eI$, $\eII$, and $\eIII$.
At the Fermi level the conductivities are $\sigma_{xx} = \sigma_{yy} \approx 8.5 \times 10^{5}~\Omega^{-1}\,\mathrm{m}^{-1}$ and $\sigma_{zz} \approx7.5 \times 10^{5}~\Omega^{-1}\,\mathrm{m}^{-1}$.
}
\end{figure}
First, to demonstrate the accuracy of our calculation method, we show the chemical potential dependence of the electrical conductivity components, $\sigma_{i j}$ for $i, j = x, y, z$, in Fig.~\ref{fig:electrical_conductivity_tensor}.
The upper panels in Fig.~\ref{fig:electrical_conductivity_tensor} depicts the longitudinal conductivity components, $\sigma_{i i}$ ($i = x, y, z$), for the $\hole$ valley, the sum of $e$ valleys, and the total of the valleys.
The corresponding energy dispersions are shown in Fig.~\ref{fig:bz-fs}~(d) and (e).
The total longitudinal conductivity tensor is less anisotropic at the Fermi level $\mu = 0$, though the contribution from one electron pocket is highly anisotropic.

The transverse conductivity components are shown in the lower panels in Fig.~\ref{fig:electrical_conductivity_tensor}, and all the components are zero within the margin of error.
As shown in SM, for each $e$ valley, $\sigma_{x x}$ and $\sigma_{y y}$ are not equal, but the sums of $e$ valleys take the same value.
Moreover, the transverse components of each $e$ valley take nonzero values, although the transverse components of the sum of $e$ valleys are zero.
These anisotropies are consistent with the crystalline symmetry, which indicates the high accuracy of our calculations.

\begin{figure}
\centering
\includegraphics[width=\linewidth,clip]{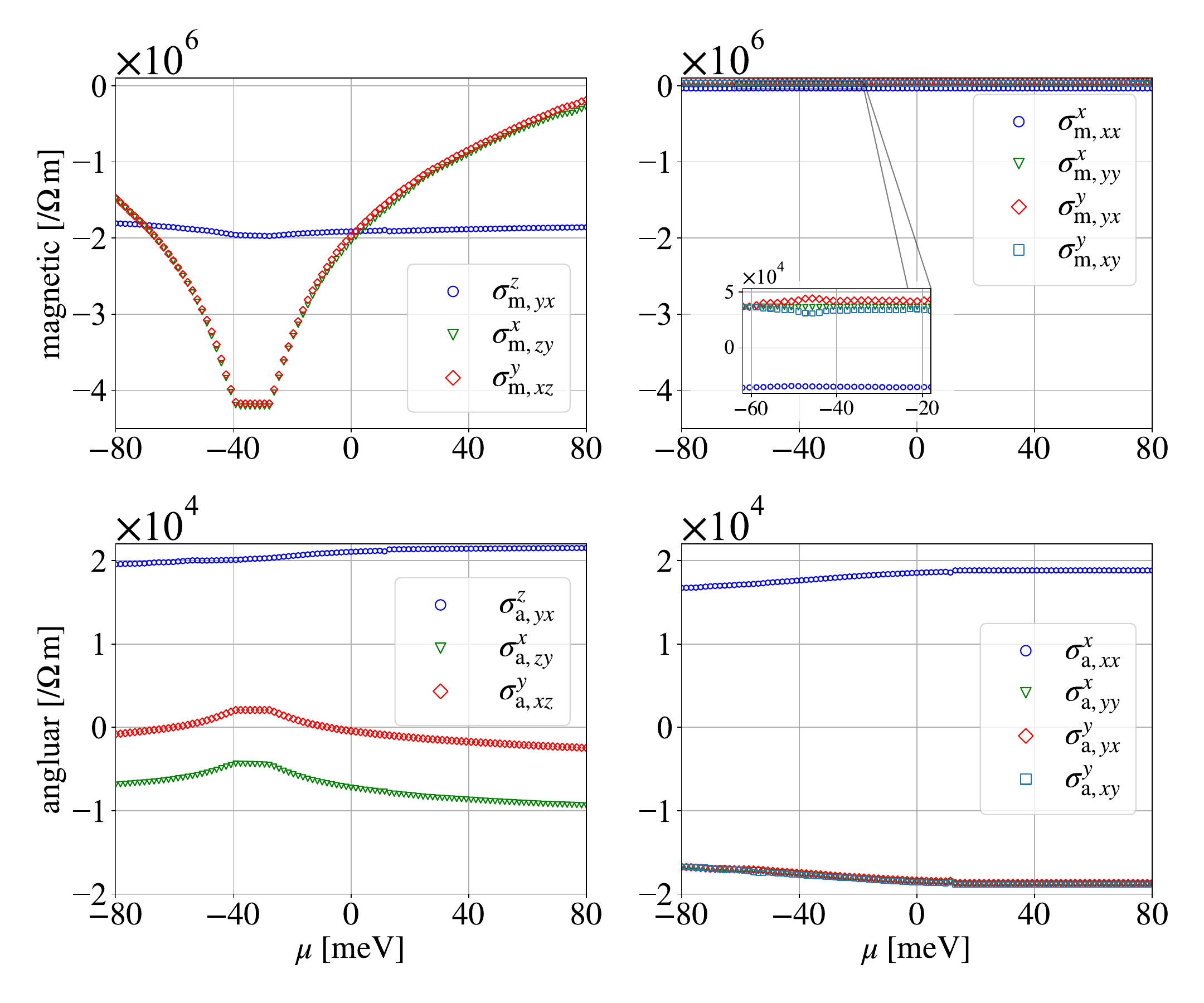}
\caption{\label{fig:spin_conductivity_tensor}
Chemical-potential dependence of representative components of the total spin Hall-conductivity (SHC) tensor obtained from (upper panels) the SMM picture and (lower panels) the SAM picture.
Whereas the SAM-based SHC is strongly anisotropic, the SMM-based counterpart is markedly less so.
Unusual components, e.g. $\sigma_{xx}^{x}$ (with $q =\mathrm{m}$ for SMM and $q = \mathrm{a}$ for SAM), acquire finite values that depend on the damping constants.
The damping rates for the $\hole$ pocket and the $\eI$, $\eII$, and $\eIII$ pockets are identical to those used in Fig.~\ref{fig:electrical_conductivity_tensor}.
Chemical-potential dependences of the remaining tensor components are provided in SM.}
\end{figure}
Figure~\ref{fig:spin_conductivity_tensor} depicts the chemical potential dependence of the spin Hall conductivity tensors, $\sigma_{\mathrm{m}, i j}^{\alpha}$ and $\sigma_{\mathrm{a}, i j}^{\alpha}$.
Left panels of Fig.~\ref{fig:spin_conductivity_tensor} show the transverse components of SHCs, such as $\sigma_{q, y x}^z$ ($q = \mathrm{m}, \mathrm{a}$).
The magnitudes of the conductivities are quite different: $\sigma_{\mathrm{m}, i j}^{\alpha}$ are two orders of magnitude larger than $\sigma_{\mathrm{a}, i j}^{\alpha}$.
Moreover, SMM-based SHC has the characteristic isotropy $\sigma_{\mathrm{m}, zy}^x = \sigma_{\mathrm{m}, xz}^y \simeq \sigma_{\mathrm{m}, y x}^z$ at the Fermi level.
In contrast, such a character is never seen in $\sigma_{\mathrm{a}, i j}^{\alpha}$, and the conductivity tensor is highly anisotropic as discussed in the previous works~\cite{guo2022,qu2023}.
Note that the SMM-based SHC can be anisotropic by changing the chemical potential (doping electrons or holes).
Although the absolute values of SHCs depend on the energy cutoff, the anisotropies of SHCs remain, and the magnitudes do not change significantly by changing $|E_{\mathrm{cutoff}} - E_{\mathrm{F}}| = 100~\mathrm{meV}$~(see SM for details).
We also confirm that the transverse components of SHCs do not depend on the damping constants.
The recent spin-torque experiments~\cite{chi2022} have shown the carrier concentration dependence of the spin Hall conductivity, which does not conflict with our SMM-based SHC.

The right panels of Fig.~\ref{fig:spin_conductivity_tensor} show the longitudinal components such as $\sigma_{q, xx}^x$ ($q = \mathrm{m}, \mathrm{a}$), which have non-zero values, permitted by crystalline symmetry~\cite{guo2022}.
For SAM-based SHC, the longitudinal components are in the same order as the transverse components, such as $\sigma_{\mathrm{a}, y x}^z$.
In contrast, the longitudinal components of the SMM-based SHC are much less than the transverse components.
We find that the longitudinal components are proportional to the damping constants, and hence, their origin differs from that of transverse components based on the spin Berry curvature (see SM for more details).

Here, we discuss the spin Hall angles.
We define the spin Hall angles as $\theta_{q, j i}^{\alpha} = \tilde{\sigma}_{q, j i}^{\alpha} / \sigma_{i i}$ with $q = \mathrm{m}, \mathrm{a}$ and evaluate them as
\begin{align}
& \theta_{\mathrm{m}, y x}^z
    \simeq - 2.2
, \quad
\theta_{\mathrm{m}, z y}^x
    \simeq - 2.2
, \quad
\theta_{\mathrm{m}, x z}^y
    \simeq - 2.5
, \\
& \theta_{\mathrm{a}, y x}^z
    \simeq 0.24
, \quad
\theta_{\mathrm{a}, z y}^x
    \simeq - 0.082
, \quad
\theta_{\mathrm{a}, x z}^y
    \simeq - 0.0057
\label{eq:SHA_SAM}
.\end{align}
The SMM-based spin Hall angles are negative and their amplitudes exceed unity, while the SAM-based ones take various values.

Here, we give comments on the previous experimental works~\cite{fukumoto2023,emoto2016,yue2018,shen2021}.
We evaluate SHCs normal to the $(110)$ and $(111)$ planes in textured polycrystalline films, which are observed in the recent spin-torque experiments~\cite{fukumoto2023}.
The evaluation procedures are shown in SM, and the results are obtained as
\begin{align*}
\sigma_{\mathrm{m}}^{(110)} = -1.9 \times 10^6~\mathrm{/\Omega \, m}
, & \quad
\sigma_{\mathrm{m}}^{(111)} = -2.0 \times 10^6~\mathrm{/\Omega \, m}
, \\
\sigma_{\mathrm{a}}^{(110)} = 4.5 \times 10^3~\mathrm{/\Omega \, m}
, & \quad
\sigma_{\mathrm{a}}^{(111)} = -3.4 \times 10^2~\mathrm{/\Omega \, m}
.\end{align*}
These results are not inconsistent with experimental indications; however, we believe it is still premature to make a quantitative comparison.
This is because the interplay between the flows of SAM and SMM remains an open problem.
Furthermore, a comparison with experimental results obtained from inverse spin Hall effect measurements~\cite{emoto2016,yue2018,shen2021} is also not yet feasible.
Although the conversion of spin current into charge current in bismuth can, in principle, be described by the SHC via Onsager reciprocity, the spin current injected from a ferromagnet may contain both SAM and SMM components.
How these components are converted into charge current remains to be clarified in future studies.

We finally discuss previous theoretical works on the SHC based on SAM in bismuth~\cite{sahin2015,guo2022,qu2023}.
In the work by \c{S}ahin and Flatt\'{e}~\cite{sahin2015}, using a third-nearest-neighbour $sp^{3}$ tight–binding model for $\mathrm{Bi}_{1-x} \mathrm{Sb}_x$, the intrinsic spin Hall conductivity was computed from the Kubo-Berry formalism.
At zero doping the values are $\sigma^{z}_{\mathrm{a}, yx} \simeq 4.74 \times 10^4 (\hbar/e)~\mathrm{/\Omega\,m^{-1}}$ for Bi.
Guo investigated three independent tensor components of both spin Hall and spin Nernst conductivities in rhombohedral Bi based on fully relativistic DFT calculations~\cite{guo2022}.
The dominant component reaches $\sigma^{z}_{\mathrm{a}, yx} \simeq 1.06 \times 10^5~(\hbar/e)\mathrm{/\Omega\,m^{-1}}$, whereas $\sigma^{y}_{\mathrm{a}, xz}$ and $\sigma^{x}_{\mathrm{a}, zy}$ differ by $\sim 20 \%$, evidencing pronounced anisotropy.
The longitudinal components, such as $\sigma_{\mathrm{a}, xx}^x$, were reported to be zero.
Qu and Tatara calculated the intrinsic orbital Hall conductivity and SHC of Bi~\cite{qu2023}, employing the Liu-Allen model.
They found $\sigma^{z}_{\mathrm{a}, yx} \simeq 9.5 \times 10^{4}~(\hbar/e)\mathrm{/\Omega\,m^{-1}}$, whereas the corresponding orbital Hall conductivity is roughly three times smaller.
The longitudinal components were reported to be nonzero but not discussed.
In comparison to the above, our resultant SAM-based SHCs take smaller values because of the cutoff energy.
However, our resultant SAM-based SHCs are much more anisotropic as shown in Eq.~(\ref{eq:SHA_SAM}), where such anisotropy is caused by the $\hole$ contribution (see SM for the details).
Moreover, we report for the first time that the longitudinal components depend on the damping rate, and hence, their origin differs from the spin Berry curvature.

Although our analysis focused on elemental Bi, the distinction between SAM and SMM is generic and can qualitatively reshape the interpretation of all spin Hall measurements.
In $\mathrm{W Te}_2$, giant charge-to-spin conversion has already been reported with spin-orbit torques that are greater than SAM-based estimates by an order of magnitude~\cite{zhou2019,shi2021}.
In GaAs quantum wells, where weak spin-orbit coupling is usually treated within a single-band picture, the multiband admixture responsible for SMM currents demands a re-evaluation of the spin accumulation~\cite{kato2004,wunderlich2005}.
Likewise, in three-dimensional topological insulators such as $\mathrm{Bi}_2 \mathrm{Se}_3$ or $\mathrm{(Bi, Sb)}_2 \mathrm{Te}_3$, the helical surface states carry zero SAM-based SHC but a finite SMM-based response, naively suggesting that Kerr-rotation and spin-torque experiments couple predominantly to the latter~\cite{mellnik2014,chi2020}.
A systematic re-analysis across these platforms should thus uncover overlooked SMM contributions and may reconcile several outstanding discrepancies between theory and experiment.

In conclusion, we have clarified the essential distinction between spin angular momentum (SAM) and spin magnetic moment (SMM) in the spin Hall effect of multiband systems.
Within the three-dimensional Dirac model, the intrinsic spin Hall conductivity (SHC) associated with SAM vanishes, whereas that associated with SMM remains finite.
Employing a $\bk\!\cdot\!\bp$ theory for elemental bismuth and the linear response theory, we computed the complete intrinsic SHC tensors and found that the SMM-based SHC is approximately two orders of magnitude larger and markedly less anisotropic than the SAM-based one.  
These results demonstrate that the intrinsic spin Hall response depends critically on whether the transported quantity is SAM or SMM.  
Because the underlying $\bk\!\cdot\!\bp$ theory requires only the low-energy band structure; the approach is readily extendable to narrow-gap semiconductors, Dirac electrons, and other topological materials in which multiband effects are prominent.

\begin{acknowledgments}
The authors thank M.~Shiraishi and M.~Hayashi.
This work was supported by JSPS KAKENHI Grant Numbers JP22K13997, JP22K18318, JP23H00268, and JP25K07218, by JST-CREST Grant Number JPMJCR19J4, and by JST-Mirai Program Grant Number JPMJMI20A1.
\end{acknowledgments}

\bibliography{bismuth,cgal}
\end{document}


%
%
%
%
%
\title{Supplementary Material for\\
Magnetic Moment vs Angular Momentum: Spin Hall Response in Bismuth}
\date{\today}
\author{Junji Fujimoto}
\affiliation{Department of Electrical Engineering, Electronics, and Applied Physics, Saitama University, Saitama, 338-8570, Japan}
\email[E-mail address: ]{jfujimoto@mail.saitama-u.ac.jp}

\author{Yuki Izaki}
\affiliation{Department of Physics, Tokyo Institute of Technology, 2-12-1 Ookayama, Meguro-ku, Tokyo 152-8551, Japan}

\author{Yuki Fuseya}
\affiliation{Department of Engineering Science, University of Electro-Communications, Chofu, Tokyo 182-8585, Japan}

\maketitle
\tableofcontents

\section{\label{sec:basis}Low-energy basis and symmetry at the $L$ and $T$ points}
In this section, we summarize the construction of the low-energy basis states and their symmetry properties at the $L$ and $T$ points, which underlie the matrix representations of the spin angular momentum~(SAM) and spin magnetic moment~(SMM) operators used in the main text.

Because the spin-orbit coupling in bismuth is strong, the low-energy Bloch states at the $L$ and $T$ points cannot be classified by real-spin eigenstates.
Instead, each band forms a Kramers doublet protected by time-reversal symmetry, and we introduce a Kramers (pseudospin) index $\eta=1,2$ to label the two degenerate partners.

At the $L$ point, the low-energy electronic structure is well described by retaining one conduction-like and one valence-like band, each forming a Kramers doublet.
The valence-band states originate from the $L_5$ and $L_6$ representations, which form a Kramers pair related by time-reversal symmetry and are commonly denoted as $L_s = L_5 + L_6$.
Similarly, the conduction-band states originate from the $L_7$ and $L_8$ representations, which also form a Kramers pair and are denoted as $L_a = L_7 + L_8$~\cite{golin1968,falicov1965}.
Both $L_s$ and $L_a$ are invariant under time-reversal symmetry. Under spatial inversion, $L_s$ is symmetric (even), whereas $L_a$ is antisymmetric (odd).
A convenient basis for the minimal low-energy subspace at the $L$ point is therefore
\begin{align}
|L \rangle = \{|L_a,\eta=1\rangle, |L_a,\eta=2\rangle, |L_s,\eta=1\rangle, |L_s,\eta=2\rangle\}.
\end{align}
In our formulation, the SAM operator is defined microscopically in the underlying atomic-orbital basis by assigning the real-spin Pauli matrices to the spin degree of freedom.
The $k\cdot p$ SAM operator used in the main text is obtained by projecting this microscopic SAM onto the retained low-energy subspace spanned by $|L\rangle$.
By contrast, the SMM operator is not assumed to be proportional to the SAM.
Instead, it is constructed from the general $g$-factor formula within $k\cdot p$ theory~\cite{fuseya2015}, and is then expressed in the same low-energy basis.

The electronic structure at the $T$ point cannot be reduced to a two-band (Dirac-type) description. Within the relevant energy window around the Fermi level, there are six bands, which become twelve when Kramers degeneracy is taken into account. In order of increasing energy, these bands can be labeled as $T_6^-$, $T_6^+$, $T_{45}^-$, $T_6^+$, $T_6^-$, and $T_{45}^+$~\cite{golin1968,falicov1965}.
\begin{align}
|T\rangle = \{ |T_6^-\rangle, |T_6^+\rangle, |T_{45}^-\rangle, |T_6^+\rangle, |T_6^-\rangle, |T_{45}^+\rangle \}
\end{align}
with
\begin{align}
|T_6^-\rangle & = \{ |T_6^-,\eta=1\rangle, |T_6^-,\eta=2\rangle \}
, \notag \\
|T_6^+\rangle & = \{ |T_6^+,\eta=1\rangle, |T_6^+,\eta=2\rangle \}
, \notag \\
|T_{45}^-\rangle & = \{ |T_{45}^-,\eta=1\rangle, |T_{45}^-,\eta=2\rangle \}
, \notag \\
|T_6^+\rangle & = \{ |T_6^+,\eta=1\rangle, |T_6^+,\eta=2\rangle \}
, \notag \\
|T_6^-\rangle & = \{ |T_6^-,\eta=1\rangle, |T_6^-,\eta=2\rangle \}
, \notag \\
|T_{45}^+\rangle & = \{ |T_{45}^+,\eta=1\rangle, |T_{45}^+,\eta=2\rangle \}
.\end{align}
The Fermi level crosses the $T_{45}^- = T_4^- + T_5^-$ band.
Accordingly, for the $T$-valley hole pocket we construct the effective Hamiltonian and the SAM/SMM operators within this twelve-band $k\cdot p$ subspace. This multiband treatment is essential to faithfully describe the low-energy wave functions and the projected spin operators at the $T$ point, and cannot be captured within a $4\times4$ Dirac Hamiltonian.

\section{\label{sec:def_SAM_SMM}Definition and construction of SAM and SMM operators in the $k\cdot p$ framework}
We define the spin angular momentum (SAM) operator microscopically as the intrinsic electron spin in the underlying atomic-orbital (tight-binding) basis.
Following the Liu-Allen tight-binding model~\cite{liu1995}, let $| n, s \rangle$, $|n, x \rangle$, $|n, y \rangle$, $|n, z \rangle$ denote the $6s$, $6p_x$, $6p_y$, $6p_z$ orbital basis (including the spin degree of freedom) with the atom index $n = 1, 2$.
Hence, the basis is given by
\begin{align}
|\Psi\rangle
    & = \{ |1,s\rangle, |1,x\rangle, |1,y\rangle, |1,z\rangle, |2,s\rangle, |2,x\rangle, |2,y\rangle, |2,z\rangle\}
\end{align}
The microscopic SAM operator is introduced as
\begin{align}
s^{\alpha}
    & = \frac{\hbar}{2} \begin{pmatrix}
        \sigma^{\alpha}
    &
    &
    &
    &
    &
    &
    &
    \\
    &   \sigma^{\alpha}
    &
    &
    &
    &
    &   0
    &
    \\
    &
    &   \sigma^{\alpha}
    &
    &
    &
    &
    &
    \\
    &
    &
    &   \sigma^{\alpha}
    &
    &
    &
    &
    \\
    &
    &
    &
    &   \sigma^{\alpha}
    &
    &
    &
    \\
    &
    &
    &
    &
    &   \sigma^{\alpha}
    &
    &
    \\
    &   0
    &
    &
    &
    &
    &   \sigma^{\alpha}
    &
    \\
    &
    &
    &
    &
    &
    &
    &   \sigma^{\alpha}
    \end{pmatrix}
,\end{align}
where $\sigma^{\alpha}$ with $\alpha = x, y, z$ are the Pauli matrices.
In the orbital basis, $s^{\alpha}$ is block-diagonal with respect to orbital indices and identical in each orbital block.

Let $\{|\tau, n, \eta \rangle\}$ with $n = 1, 2, \cdots, 8$ and $\eta = 1, 2$ be the Bloch eigenstates of the tight-bindin Hamiltonian at the $\tau$ valley and let $\mathcal{P}$ be the projector onto the low-energy subspace retained for the $\tau$ valley,
\begin{align}
\mathcal{P}_{\tau}
    & = \sum_{n, \eta} \ket{\tau, n, \eta} \bra{\tau, n, \eta}
.\end{align}
The effective SAM operator used in the main text is then defined by
\begin{align}
s^{(\tau), \alpha}
    & = \mathcal{P}_{\tau} s^{\alpha} \mathcal{P}_{\tau}
.\end{align}
Their matrix representations follow once a specific low-energy basis is chosen.

At the $\tau = \eI, \eII, \eIII$ valleys, we retain one conduction-like and one valence-like band, each forming a Kramers doublet.
Using the symmetry-adapted notation of Refs.~\cite{golin1968,falicov1965}, the valence Kramers pair is denoted $L_s=L_5+L_6$ and the conduction Kramers pair is denoted $L_a=L_7+L_8$ as mentioned in Sec.~\ref{sec:basis}.

At the $\tau = \hole$ valley, several bands lie within the energy window relevant to the hole pocket. In the present calculation we retain six bands, which become twelve including Kramers degeneracy. In increasing-energy order they are labeled as $T_6^-$, $T_6^+$, $T_{45}^-$, $T_6^+$, $T_6^-$, and $T_{45}^+$, where $\pm$ denotes inversion parity, and the Fermi level crosses $T_{45}^-=T_4^-+T_5^-$. We introduce a Kramers index $\eta=1,2$ for each band and construct the effective operators within this twelve-band subspace.

On the other hand, the spin magnetic moment~(SMM) operator is defined as the coefficient of the Zeeman coupling to an external magnetic field $\bm{B}$,
\begin{align}
\mathcal{H}_{\mathrm{Zeeman}}
    & = - \bm{m} \cdot \bm{B}
.\end{align}
In a general multiband crystal with spin-orbit coupling, $\bm{m}$ is not proportional to $\bm{s}$, because it contains interband contributions induced by spin-orbit coupling and orbital mixing.
In our calculation, $\bm{m}$ is constructed within $k\!\cdot\!p$ theory from the general $g$-factor formula derived via L\"{o}wdin partitioning~\cite{fuseya2015}.
Concretely, by introducing the Kramers state as
\begin{align}
\ket{\tau, n}
    & = \{ \ket{\tau, n, \eta = 1}, \ket{\tau, n, \eta = 2} \}
\end{align}
the effective Zeeman coupling in a chosen low-energy subspace can be written as
\begin{align}
\bra{\tau, l} \bm{m} \ket{\tau, l}
    & = \sum_{n \neq l} \frac{\zi m_e \mu_{\mathrm{B}}}{E_l - E_n}
    \begin{pmatrix}
        - (\bm{t}_{ln}^{\phantom{*}} \times \bm{t}^{*}_{ln} + \bm{u}^{\phantom{*}}_{ln} \times \bm{u}_{ln}^{*})
    &   2 \bm{t}^{\phantom{*}}_{ln} \times \bm{u}^{\phantom{*}}_{ln}
    \\  - 2 \bm{t}^{*}_{ln} \times \bm{u}^{*}_{ln}
    &   \bm{t}^{\phantom{*}}_{ln} \times \bm{t}^{*}_{ln} + \bm{u}^{\phantom{*}}_{ln} \times \bm{u}_{ln}^{*}
    \end{pmatrix}
,\end{align}
where $\bm{t}^{*}_{ln}$ and $\bm{u}^{*}_{ln}$ are the complex conjugates of $\bm{t}_{ln}$ and $\bm{u}_{ln}$, and $\bm{t}_{ln}$ and $\bm{u}_{ln}$ are defined by
\begin{align}
\bm{t}_{ln}
    & = \left\langle \tau, l, \eta = 1 \left| \frac{1}{\hbar} \frac{\partial \mathcal{H}}{\partial \bk} \right| \tau, n, \eta = 1 \right\rangle
, \\
\bm{u}_{ln}
    & = \left\langle \tau, l, \eta = 1 \left| \frac{1}{\hbar} \frac{\partial \mathcal{H}}{\partial \bk} \right| \tau, n, \eta = 2 \right\rangle
.\end{align}
Here, $\mathcal{H}$ is the tight-binding Hamiltonian.

\section{\label{sec:valley-resolved_sigma_e}Valley-resolved electric conductivity}

\begin{figure}[t]
\centering
\includegraphics[width=\linewidth]{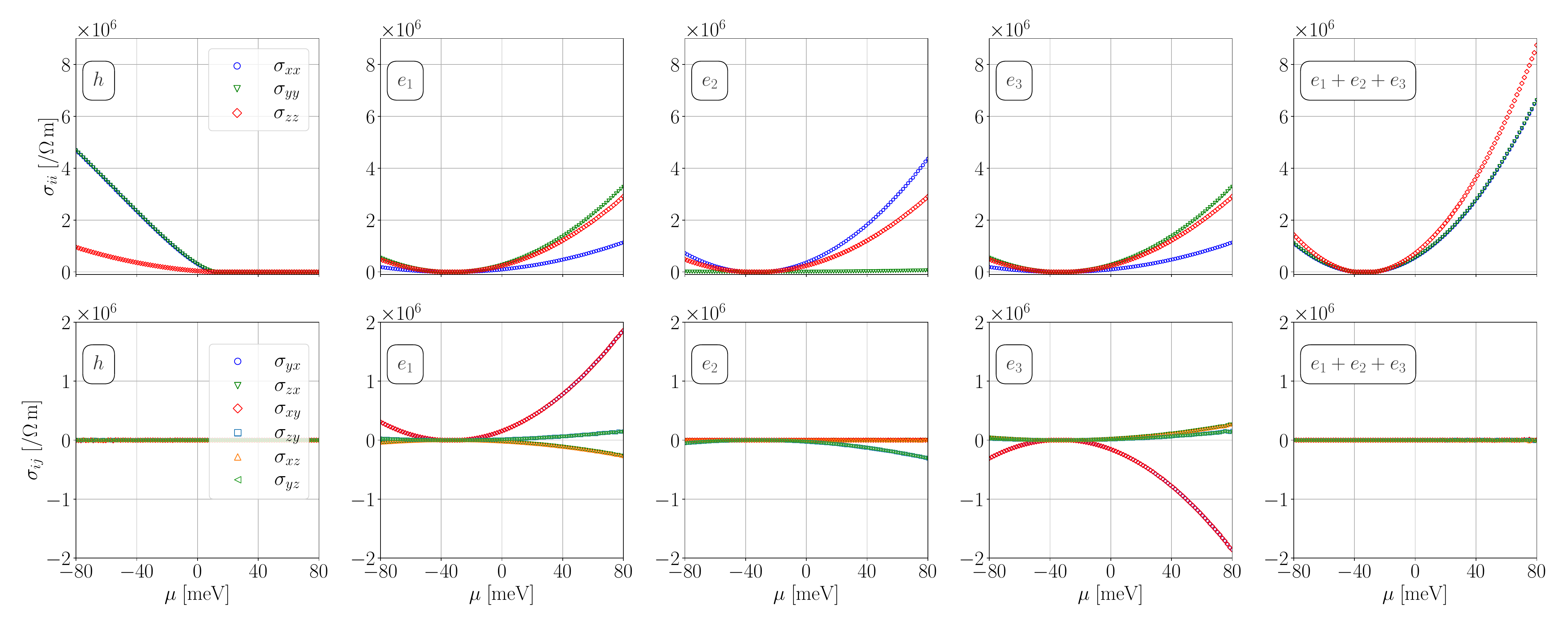}
\caption{\label{fig:sigma_e}
Calculated conductivity tensors for each valley.
The longitudinal components are finite for all valleys, and the off-diagonal components vanish for $\hole$ and the sum of $\eI, \eII, \eIII$, satisfying the full crystal symmetry individually.
These results confirm the high precision of our calculations in preserving the crystal symmetry.}
\end{figure}

We present the calculated conductivity tensors for each valley to validate the accuracy of our computations.
For the hole pocket at the $T$ point, the conductivity tensor is given by
\begin{align}
\sigma^{(\hole)}
    & = \begin{pmatrix}
            \sigma_{xx}^{(\hole)}
        &   0
        &   0
        \\  0
        &   \sigma_{yy}^{(\hole)}
        &   0
        \\  0
        &   0
        &   \sigma_{zz}^{(\hole)}
\end{pmatrix}
.\end{align}
The longitudinal components $\sigma_{ii}^{(\hole)}$ ($i = x, y, z$) are finite, while the off-diagonal components vanish.
This tensor satisfies the complete crystal symmetry on its own, reflecting the crystalline symmetry (see Fig.~\ref{fig:sigma_e}).

In contrast, each electron pocket at the three equivalent $L$ points exhibits an individual conductivity tensor
\begin{align}
\sigma^{(e_i)}
    & = \begin{pmatrix}
            \sigma_{xx}^{(e_i)} & \sigma_{xy}^{(e_i)} & \sigma_{xz}^{(e_i)}
        \\  \sigma_{yx}^{(e_i)} & \sigma_{yy}^{(e_i)} & \sigma_{yz}^{(e_i)}
        \\  \sigma_{zx}^{(e_i)} & \sigma_{zy}^{(e_i)} & \sigma_{zz}^{(e_i)}
    \end{pmatrix}
,\end{align}
where off-diagonal components are nonzero.
Each $\sigma^{(e_i)}$ individually breaks the crystal symmetry.
However, when summed over the three electron pockets, the total contribution $\sum_{i=1}^{3} \sigma^{(e_i)}$ restores the complete crystal symmetry, as required (see Fig.~\ref{fig:sigma_e}).

This analysis confirms that our computational method preserves the symmetry properties with high precision, ensuring the reliability of the calculated transport coefficients.

\section{\label{sec:valley-resolved_SHC}Spin conductivity tensors}

\begin{figure}
\centering
\includegraphics[width=0.75\linewidth]{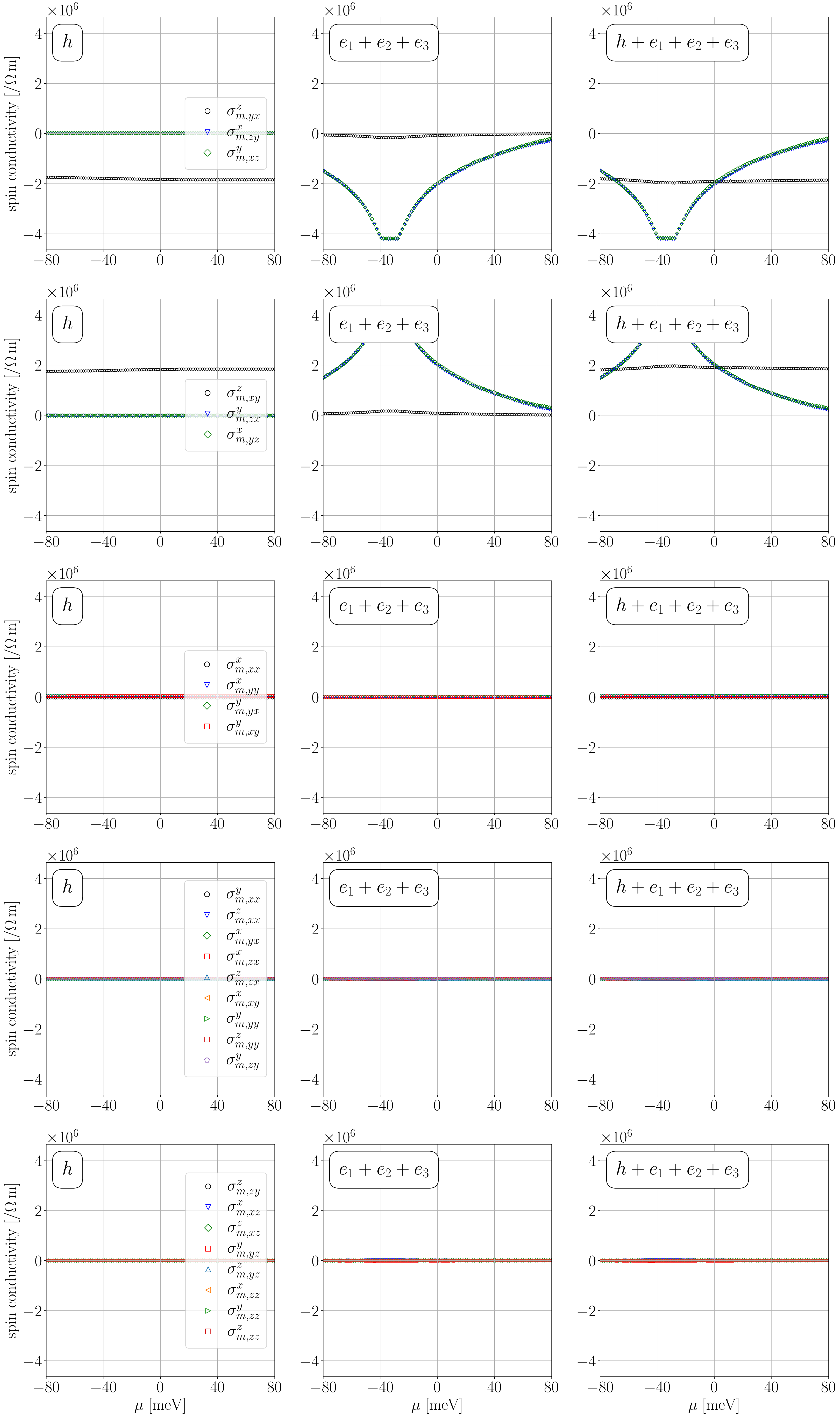}
\caption{\label{fig:sigma_m}Chemical potential dependences of the SMM-based spin conductivity tensor.}
\end{figure}

\begin{figure}
\centering
\includegraphics[width=0.75\linewidth]{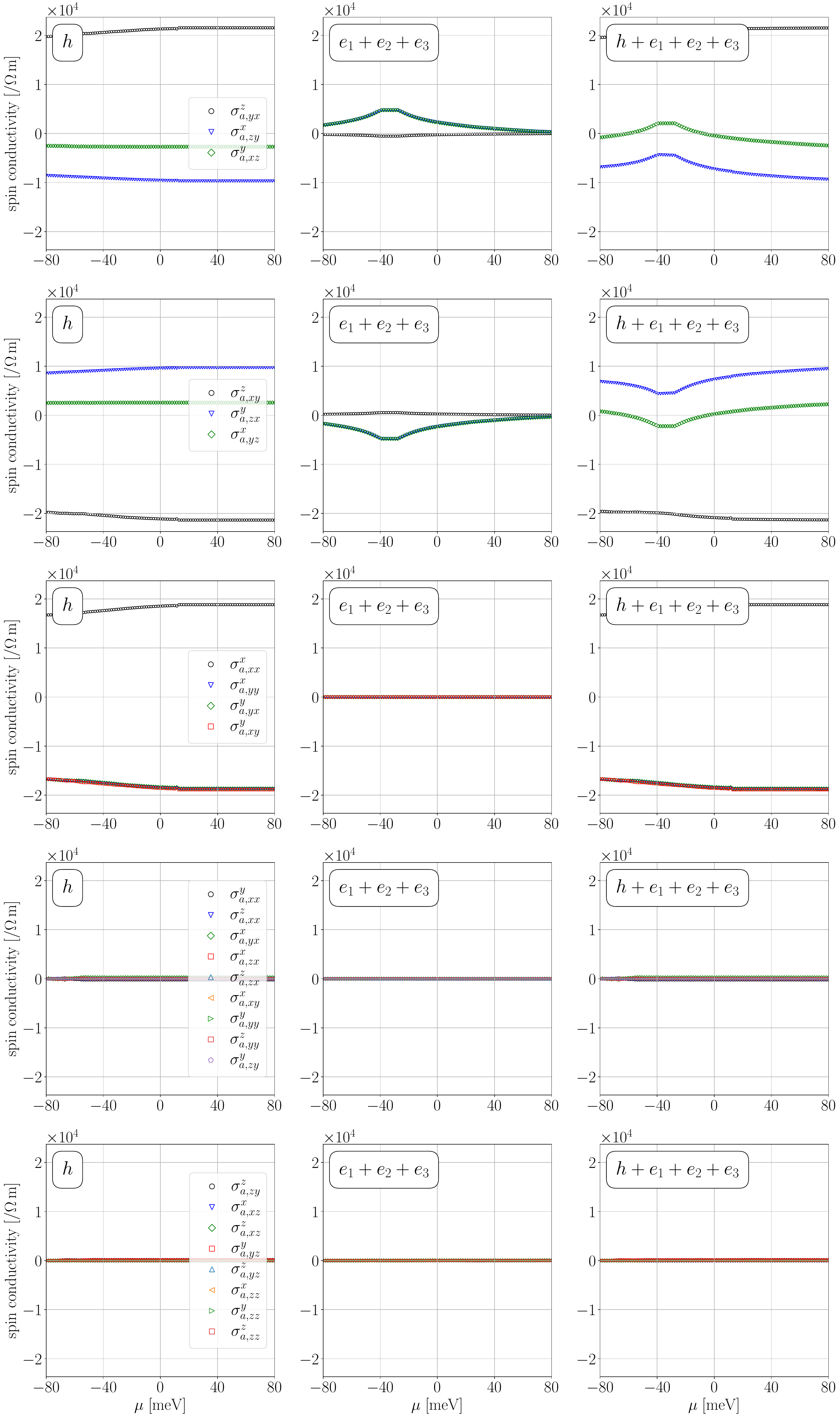}
\caption{\label{fig:sigma_a}Chemical potential dependences of the SAM-based spin conductivity tensor.}
\end{figure}

Figures~\ref{fig:sigma_m} and \ref{fig:sigma_a} show all components of the spin conductivity tensors.
We confirm that the tensors satisfy the required crystalline symmetry.

At the $\eI, \eII, \eIII$ valleys, the band gap is relatively small, and the saturation of the spin Hall conductivity inside the gap is clearly visible in Fig.~\ref{fig:sigma_m} and \ref{fig:sigma_a}.
In contrast, the band gap at the $\hole$ valley is much larger.
Within the chemical-potential range shown, the calculation probes mainly the valence-band side at the $T$ point, where the conductivity varies only weakly with chemical potential.
As a result, the saturation of the conductivity upon entering the gap is not apparent on the scale of the figure.
The conduction-band side at the $\hole$ valley lies outside the validity range of the present $k\!\cdot\!p$ description and is therefore not included.

\section{\label{sec:longitudinal_components}Symmetry and origin of longitudinal spin-current components}
In this section, we discuss the symmetry properties and physical origin of the longitudinal (unusual) spin-current components that appear in the spin Hall response.

We define the spin-current operators associated with the spin magnetic moment (SMM) and spin angular momentum (SAM) as
$j_{m,i}^{\alpha}=\frac{1}{2}\{v_i, m^{\alpha}\}$, $j_{a,i}^{\alpha}=\frac{1}{2}\{v_i, s^{\alpha}\}$, where $\hat v_i$ is the velocity operator and $\hat m^{\alpha}$ ($\hat s^{\alpha}$) denotes the SMM (SAM) operator.
In the trigonal crystal symmetry of bismuth, longitudinal spin-current responses, such as $\sigma^{\alpha}_{q,ii}$ ($q=m,a$), are not forbidden by point-group symmetry, although they are not symmetry-enforced either.
Their presence is therefore allowed on symmetry grounds.
Regarding time-reversal symmetry, both the velocity operator $v_i$ and the spin-related operators $m^{\alpha}$ and $s^{\alpha}$ are odd under time reversal.
As a result, the corresponding spin-current operators $j_{q,i}^{\alpha}$ are even under time reversal.
Consequently, longitudinal spin currents are allowed in time-reversal-symmetric systems.

Our calculations show that the longitudinal spin-current components exhibit a Drude-like dependence on the scattering time.
Within the constant-damping approximation used in this work, these components scale as $\propto \tau$ (equivalently, $\propto 1/\gamma$) and therefore diverge in the formal clean limit $\gamma \to 0$.
This divergence reflects the dc treatment of a current-like, dissipative response within a constant-$\gamma$ approximation. In more realistic systems, it is expected to be regularized by finite momentum relaxation, finite measurement frequency, and/or more complete disorder treatments beyond the present approximation.

The behavior described above is qualitatively different from that of the conventional transverse spin Hall conductivity originating from the spin Berry curvature.
The latter remains finite and essentially insensitive to the damping constant $\gamma$ in the clean limit.
The longitudinal components discussed here therefore do not originate from the intrinsic spin Berry curvature.
Instead, they represent relaxation-driven, current-like responses.
They also differ qualitatively from skew scattering, as they already appear at the bubble-diagram level without vertex corrections.

\section{\label{sec:valley-resolved_SHC}Valley-resolved SHCs}
\begin{figure*}
\centering
\includegraphics[width=\linewidth]{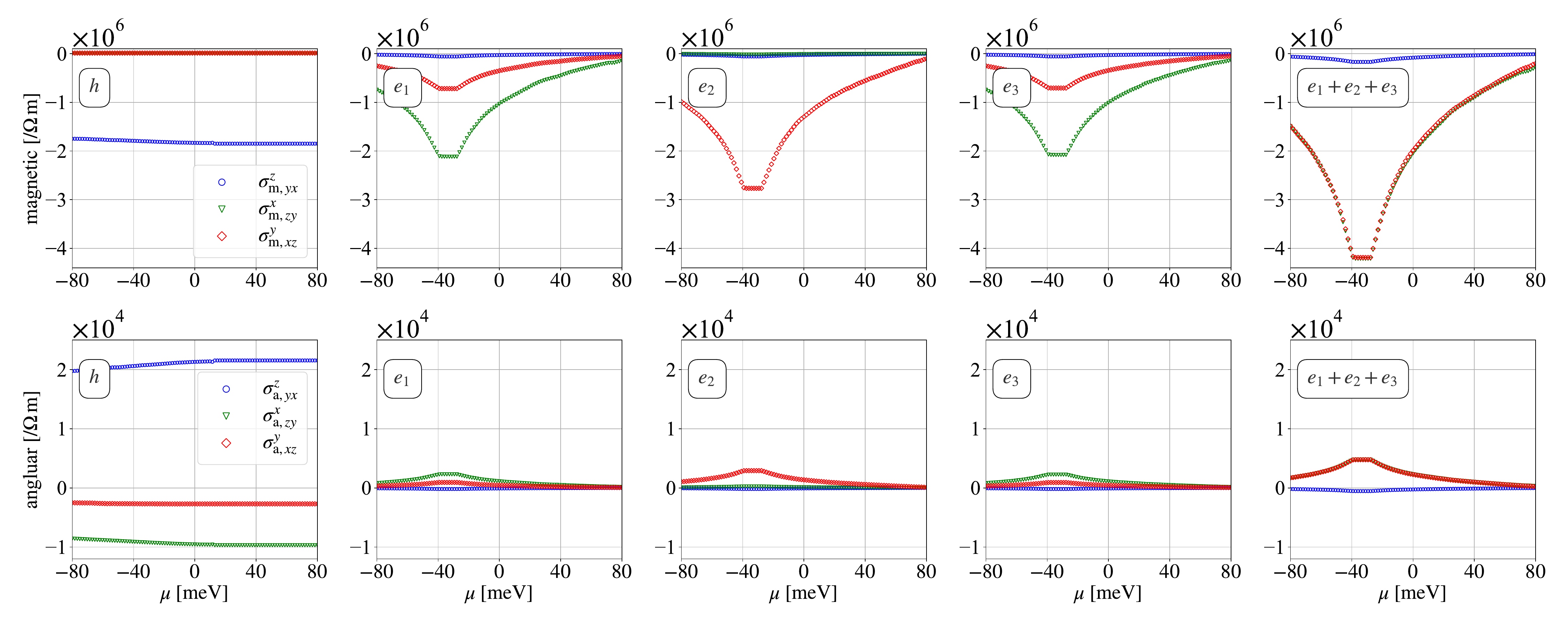}
\caption{\label{fig:valley-resolved_SHC}
Chemical-potential dependence of the valley-resolved spin Hall conductivities (SHCs) obtained from (upper panels) the SMM picture and (lower panels) the SAM picture.
The $\hole$ contributions exhibit a weak $\mu$ dependence yet are highly anisotropic, whereas the electron pockets ($e$) display the characteristic peak- or dip-like structures in both SHCs.
The SMM-based SHC is consistent with Refs.~\cite{fuseya2012,fukazawa2017}.
}
\end{figure*}
We have analyzed the valley-resolved SHCs, as shown in Fig.~\ref{fig:valley-resolved_SHC}. The SMM-based SHC exhibits relatively weak dependence on the chemical potential $\mu$ and displays nearly opposite anisotropy between $\hole$ and the sum of $\eI, \eII, \eIII$, consistent with the reduced anisotropy found in the total SHC tensor.
In contrast, the SAM-based SHC shows pronounced anisotropic features, especially in the $\hole$ contributions, which are nearly independent of $\mu$ but highly anisotropic.
The electron pockets ($\eI, \eII, \eIII$) display characteristic peak and dip structures in both the SMM- and SAM-based SHCs, reflecting the multiband nature in bismuth.
These findings emphasize that the SAM and SMM contributions arise from fundamentally different mechanisms, highlighting the importance of considering spin transport for an accurate interpretation of experimental observations.

\section{\label{sec:cutoff}Cutoff dependences of SHCs}

\begin{figure}[htbp]
\centering
\includegraphics[width=\linewidth]{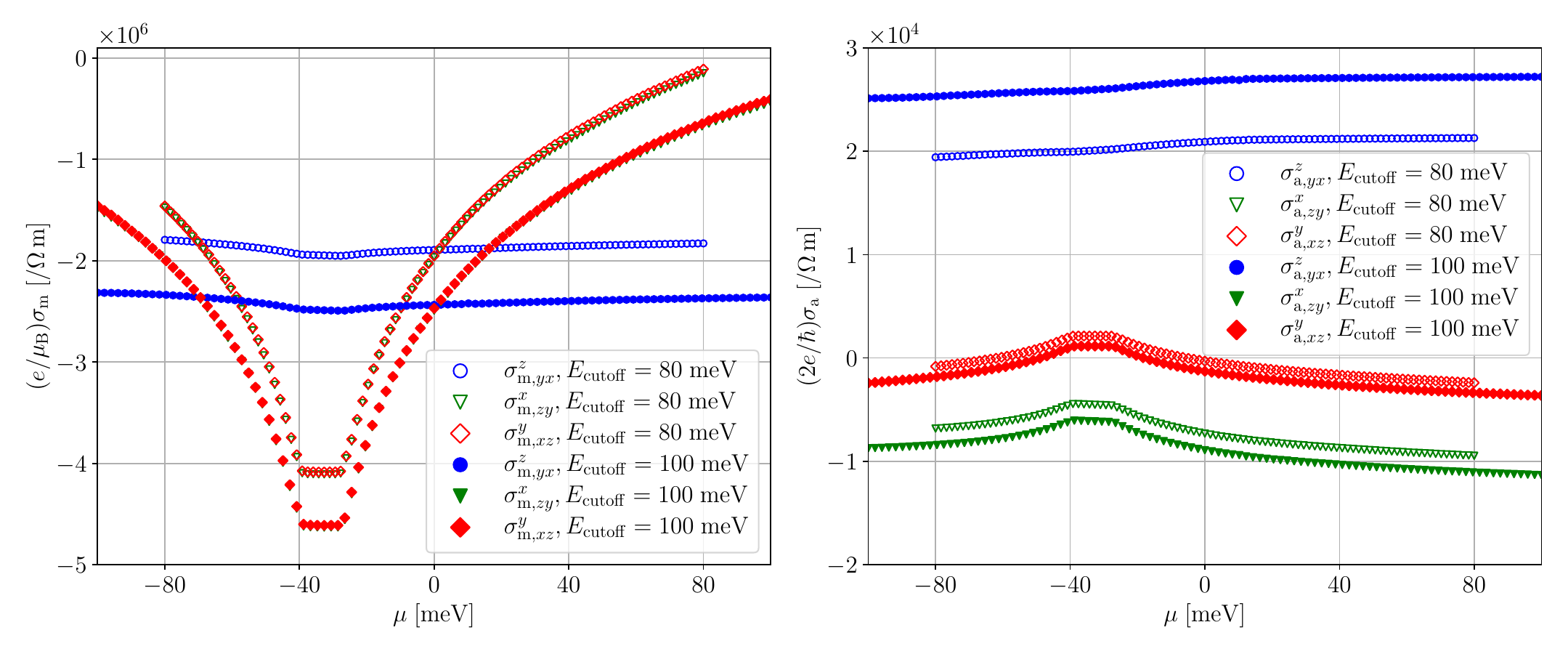}
\caption{\label{fig:cutoff}Cutoff dependence of the SHCs.
The results for cutoff energies of $80\,\mathrm{meV}$ and $100~\mathrm{meV}$ are compared.}
\end{figure}

In the main text, we presented the spin Hall conductivity calculated with an energy cutoff of $80~\mathrm{meV}$.
To examine the cutoff dependence, we have also performed calculations with a larger cutoff of $100~\mathrm{meV}$.
The results are shown in Fig.~\ref{fig:cutoff}.
As seen in the figure, the spin Hall conductivity exhibits negligible variation between the cutoffs of $80~\mathrm{meV}$ and $100~\mathrm{meV}$.
This indicates that the contribution to the spin Hall conductivity predominantly comes from the states near the band edges, and that states at higher energies have a minor influence within this range of cutoff values.

It should be noted that, due to the nature of the $\bk\!\cdot\!\bp$ expansion, contributions from the bottom of the conduction band are not captured in the present calculation.
This limitation may account for the discrepancy between the magnitude of the SHC obtained here and those reported in previous theoretical studies.
Nevertheless, within the energy window considered, the spin Hall conductivity remains essentially unchanged, confirming the validity of our calculation with a cutoff of $80~\mathrm{meV}$.

\section{\label{sec:polycry}SHCs normal to the (110) and (111) planes in textured polycrystalline films}
We compute the spin Hall conductivity $\tilde{\sigma}_{\mathrm{m}, \perp}$ and $\tilde{\sigma}_{\mathrm{a}, \perp}$ normal to the $(110)$ and $(111)$ planes in textured polycrystalline films.
A plane-fixed orthonormal basis $(X, Y, Z)$ is introduced with $\hat{Z} \parallel (hkl)$ and $(\hat{X}, \hat{Y})$ spanning the plane.
The single-crystal spin Hall tensor $\tilde{\sigma}_{q, b a}^{c} (hkl)$ with $q = \mathrm{m}, \mathrm{a}$, obtained in the crystallographic frame, is rotated into this basis via
\begin{align}
\tilde{\sigma}_{q, b a}^{c}
    & = \mathcal{R}_{a i} (hkl) \mathcal{R}_{b j} (hkl) \mathcal{R}_{c \alpha} (hkl) \sigma_{q, j i}^{\alpha}
,\end{align}
where $\mathcal{R} (hkl)$ is the rotation matrix that aligns $\hat{Z}$ with the chosen plane normal.  

Here, we focus on the case where the spin current is perpendicular to the plane, hence $b = Z$.
Inside the film, the grain orientation is random in the azimuthal angle $\phi$, so the spin current in the grain with the angle $\phi$,
\begin{align}
j_{q, Z}^{c} (\phi)
    & = \tilde{\sigma}_{q, Z X}^c E_X (\phi) + \tilde{\sigma}_{q, Z Y}^c E_Y (\phi)
,\end{align}
where $\bm{E} (\phi) = E_0 \hat{e}_{\phi}$ with  $\hat{e}_{\phi} = (\cos \phi, \sin \phi, 0)$ and the electric field magnitude $E_0$.
By taking the inner product of the spin current vector in the spin space, $\bm{j}_{q, Z} = (j_{q, Z}^X, j_{q, Z}^Y, j_{q, Z}^Z)$, and the $\hat{e}_{\phi} \times \hat{Z} = (\sin \phi, -\cos \phi, 0)$, and averaging over the angle $\phi$, we have the spin Hall conductivity normal to the $(hkl)$ plane as the macroscopic response,
\begin{align}
\tilde{\sigma}_{q, \perp} (hkl)
    & = \int_{0}^{2 \pi} d\phi \frac{\bm{j}_{q, Z} (\phi)}{E_0} \cdot (\hat{e}_{\phi} \times \hat{Z})
.\end{align}
The SHCs in main text $\sigma_{q}^{(110)}$ and $\sigma_{q}^{(111)}$ correspond to $\tilde{\sigma}_{q, \perp} (110)$ and $\tilde{\sigma}_{q, \perp} (111)$, respectively.

\bibliography{bismuth}